\documentclass[conference]{IEEEtran}

\usepackage{amssymb}
\usepackage{ifpdf}
\usepackage{hyperref}
\usepackage{graphicx}

\usepackage[caption=false]{subfig}

\usepackage{cite}
\usepackage{mathrsfs}

\ifCLASSINFOpdf
\else
\fi
\usepackage{amsmath}

\usepackage{algorithmic}

\usepackage{array}

\begin{document}

%
%

\title{Parametric Imaging of FDG-PET Data Using Physiology and Iterative Regularization: \\Application to the Hepatic and Renal Systems}

\author{\IEEEauthorblockN{Mara Scussolini\IEEEauthorrefmark{1}, Sara Garbarino\IEEEauthorrefmark{2}, Gianmario Sambuceti\IEEEauthorrefmark{3}, Giacomo Caviglia\IEEEauthorrefmark{1} and Michele Piana\IEEEauthorrefmark{4}} \IEEEauthorblockA{\IEEEauthorrefmark{1}Dipartimento di Matematica, Universit\`a di Genova, Genova, Italy\\
Email: scussolini@dima.unige.it} 
\IEEEauthorblockA{\IEEEauthorrefmark{2}Centre for Medical Image Computing, Department of Computer Science, University College London, U.K.}
\IEEEauthorblockA{\IEEEauthorrefmark{3}Dipartimento di Medicina Nucleare, IRCCS-IST San Martino, \\ and Dipartimento di Scienze della Salute, Universit\`a di Genova, Genova, Italy}
\IEEEauthorblockA{\IEEEauthorrefmark{4}Dipartimento di Matematica, Universit\`a di Genova, and CNR-SPIN, Genova, Italy}}

\maketitle

\section{Aims}
Positron Emission Tomography (PET) \cite{Bailey} is a nuclear medicine imaging technique capable of detecting the time dependent spatial distribution of pico-molar quantities of a labelled tracer, which is diffused into a living body. FDG-PET is a PET modality in which the radiopharmaceutical [18F]Fluoro-2-deoxy-D-glucose (FDG) is used as a tracer to observe metabolic processes related to glucose consumption. In order to overcome the limits of spatial resolution in PET and to improve the quality of information achievable from PET images, one approach is to develop parametric imaging methods able of showing the tracer metabolism at a local level. Starting from the design of compartmental models appropriate to describe the tracer kinetics in a predefined physiological system, parametric imaging techniques process dynamic PET data and estimate the spatial distribution of the exchange coefficients regulating tracer flow.

Most parametric imaging methods rely on linearized compartmental models and/or provide parametric images of algebraic combinations of the kinetic coefficients, as in \cite{Tsoumpas}. Rather few methods are able to reconstruct maps of each single parameter, and most of them consider simple one- and two- compartment models, as \cite{Kamasak}. We aim at developing a new approach to parametric imaging, based on a recently introduced Regions Of Interest (ROI) kinetic modeling \cite{Garbarino_kidney,Garbarino_liver}, applicable to generic compartmental models complicated enough to describe a specific organ physiology, and able to provide reliable maps of all the kinetic coefficients involved. 

\section{Methods}

\begin{figure*}[!t]
\vspace{-0.7cm}
\centering
\subfloat[Hepatic compartmental model.]{\includegraphics[width=2.6in]{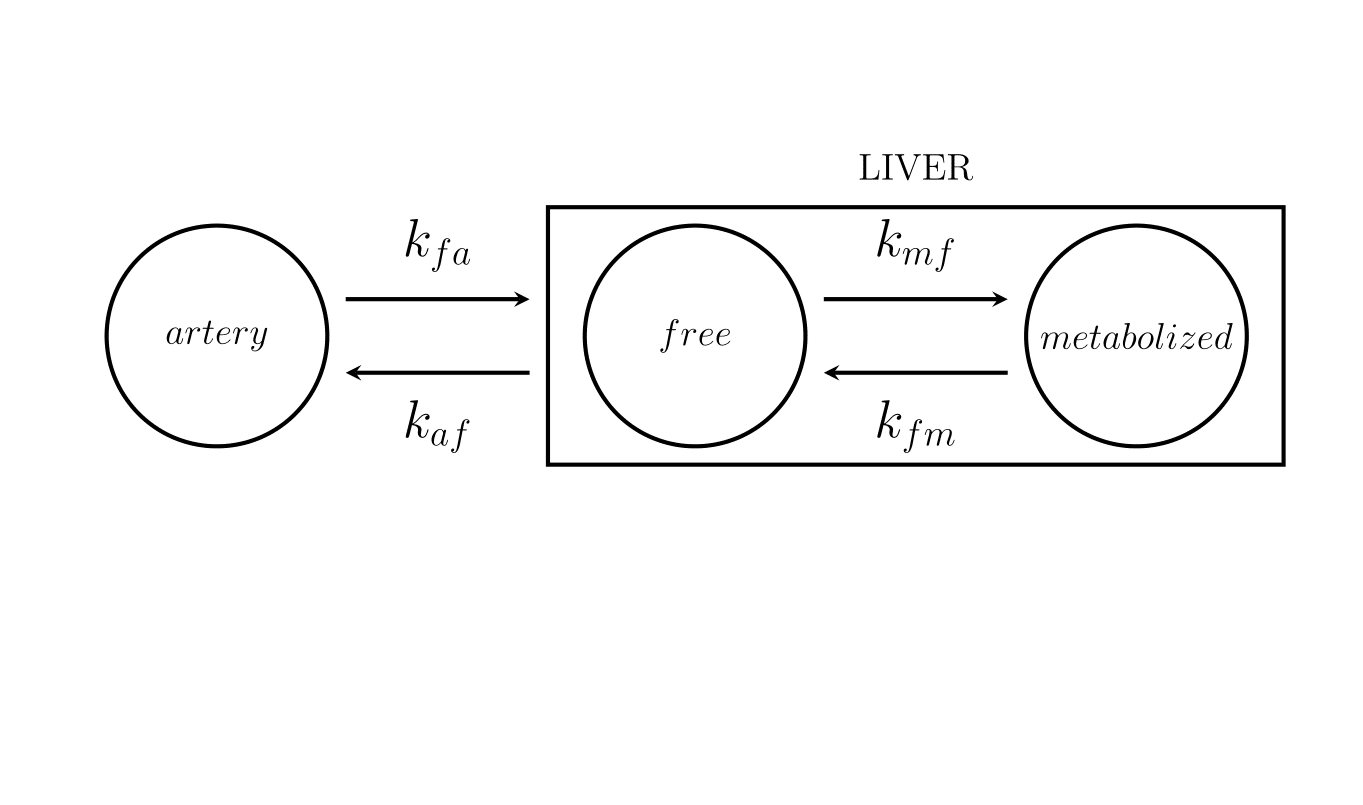} \label{fig_first_case}}
\hfil
\subfloat[Renal compartmental model. ]{\includegraphics[width=2.6in]{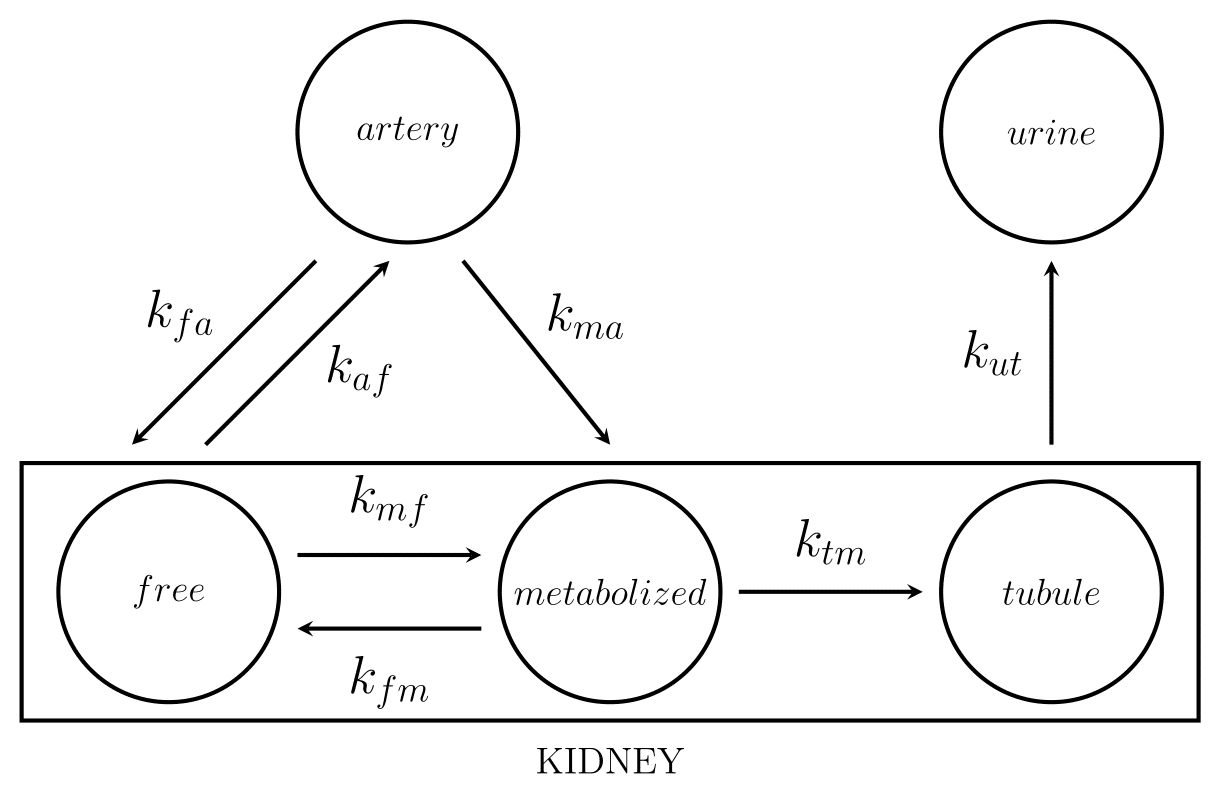} \label{fig_second_case}}
\caption{Physiology-based compartmental models.} 
\label{fig:models}
\end{figure*}

\begin{figure*}[htp]
\vspace{-0.7cm}
\centering
\subfloat[$k_{fa}$]{\includegraphics[scale=0.1]{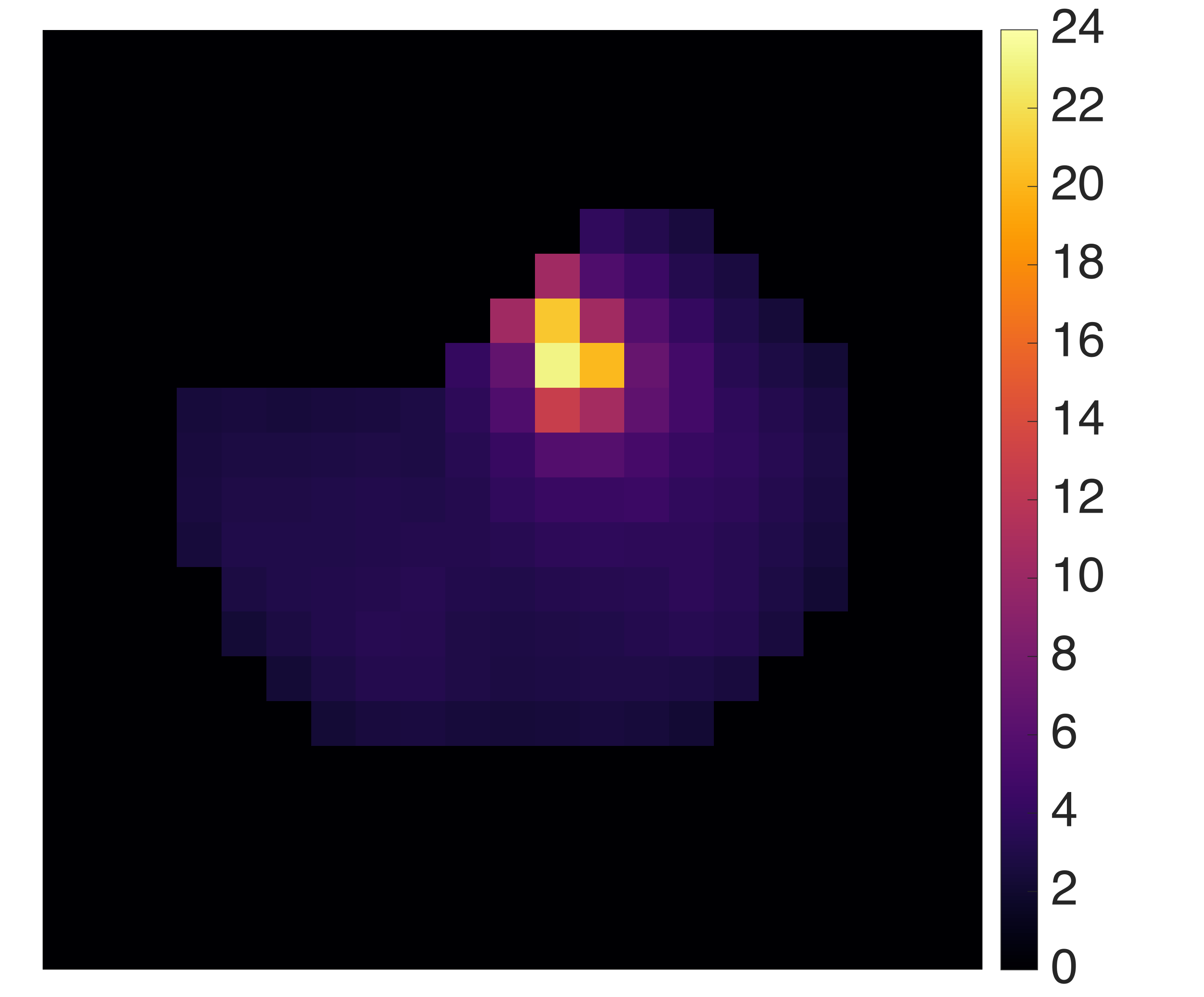} \label{fig:K1_liver}} 
\hfil
\subfloat[$k_{af}$]{\includegraphics[scale=0.1]{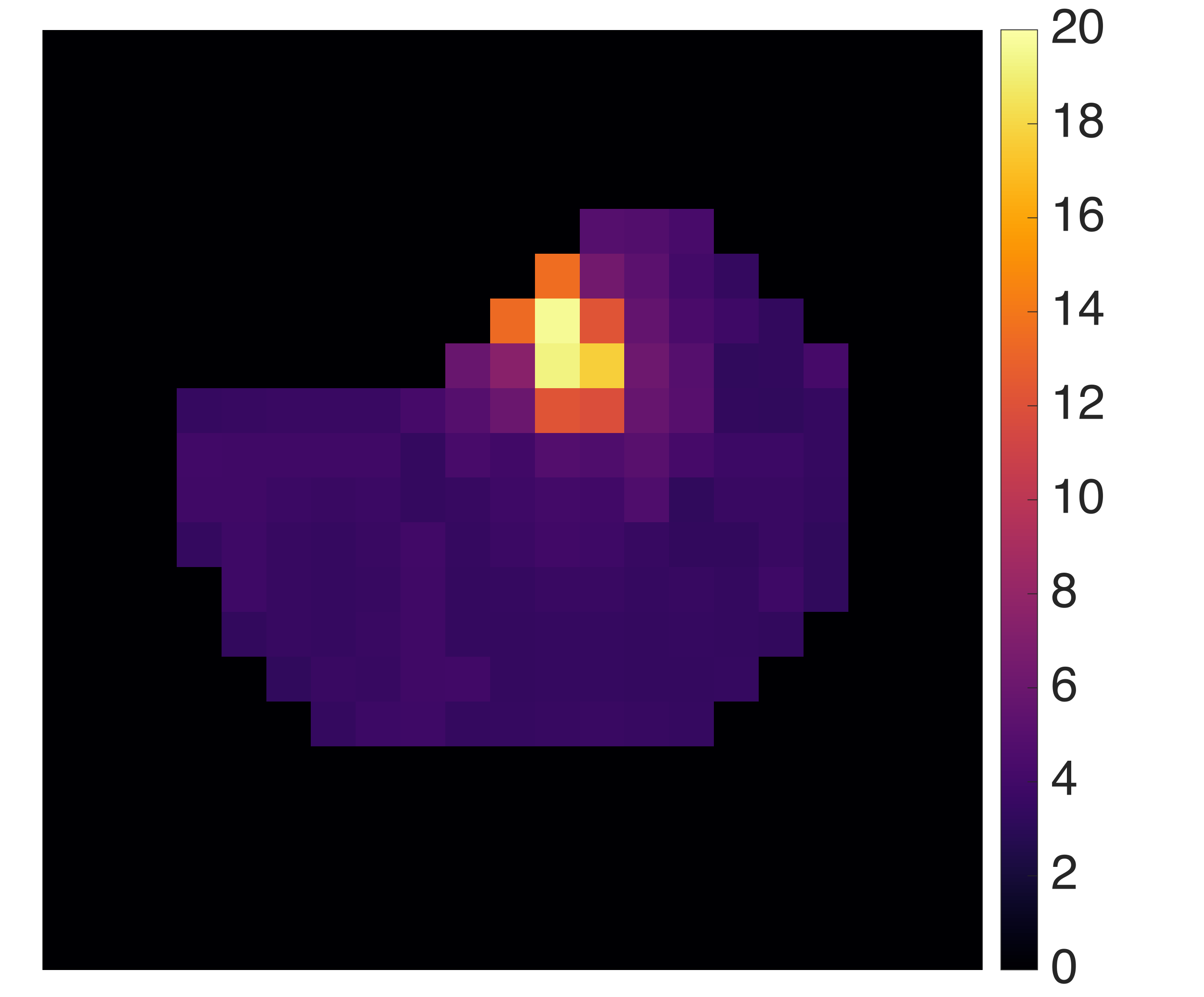} \label{fig:K2_liver}} 
\hfil
\subfloat[$k_{mf}$]{\includegraphics[scale=0.1]{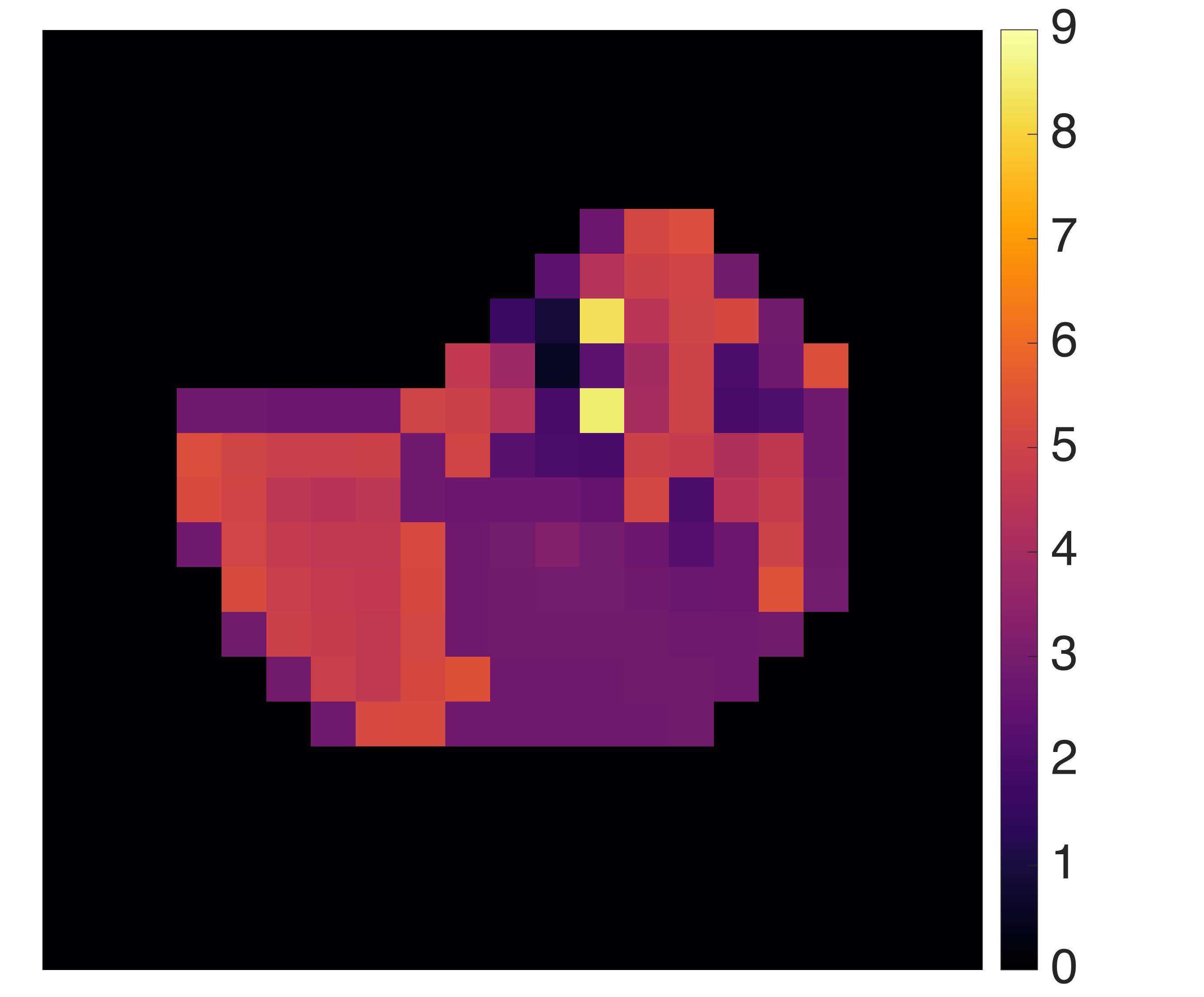} \label{fig:K3_liver}} 
\hfil
\subfloat[$k_{fm}$]{\includegraphics[scale=0.1]{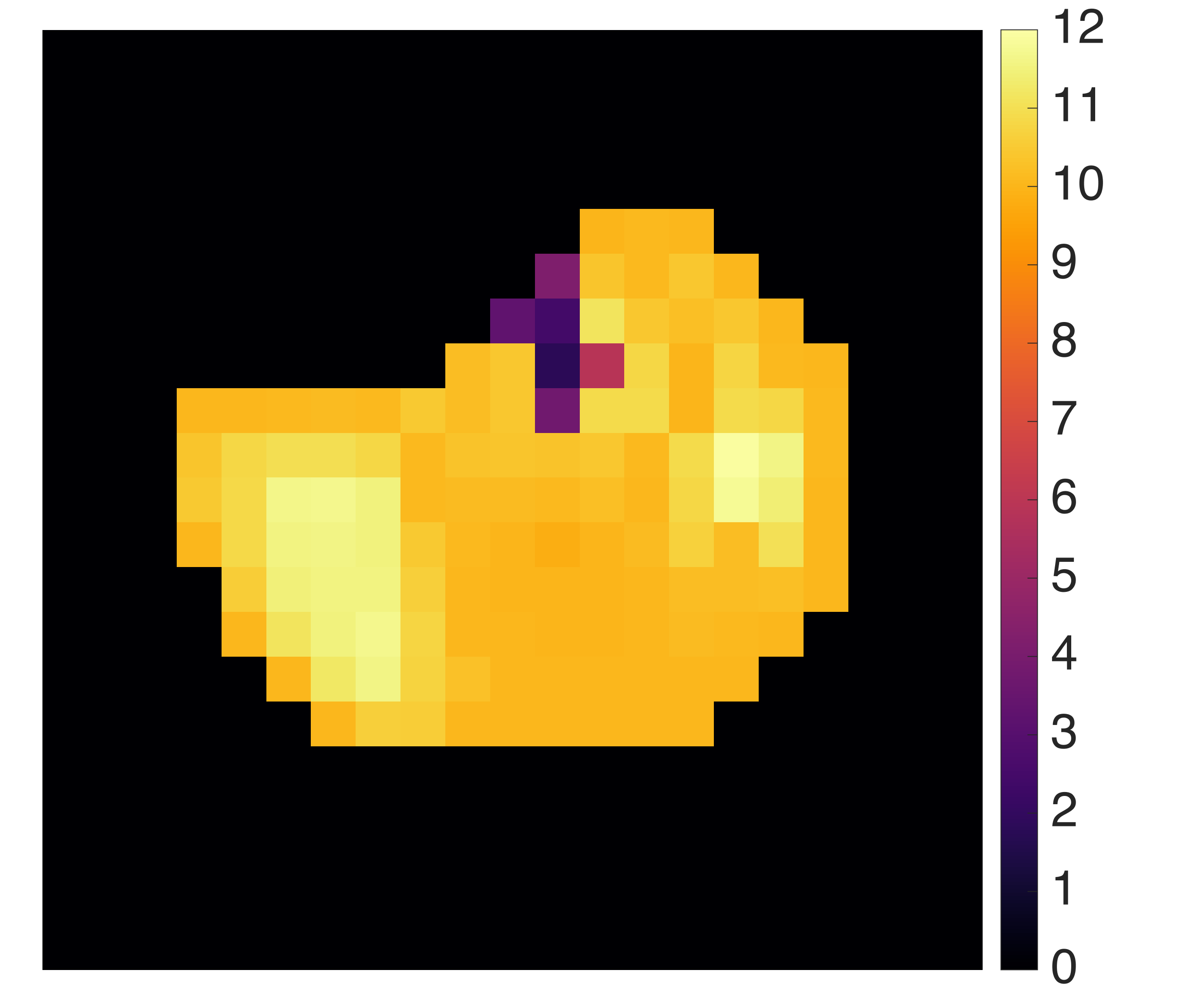} \label{fig:K4_liver}} 
\caption{Reconstructed parametric images for a selected PET slice of the hepatic system.}
\label{fig:parametric_images_liver}
\end{figure*}

\begin{figure*}[htp]
\vspace{-0.7cm}
\centering
\subfloat[$k_{fa}$]{\includegraphics[scale=0.1]{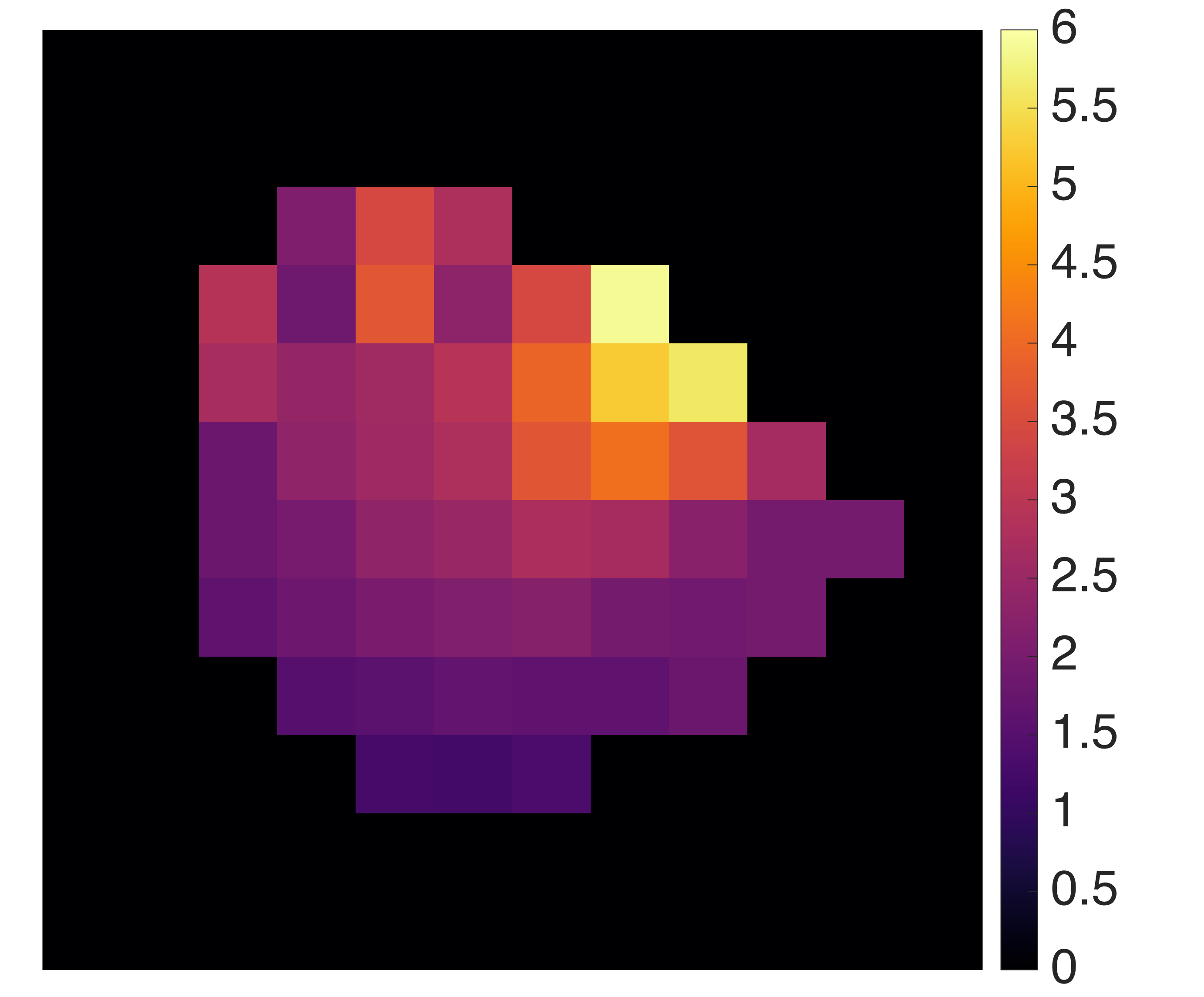} \label{fig:K1_kid}} 
\hfil
\subfloat[$k_{ma}$]{\includegraphics[scale=0.1]{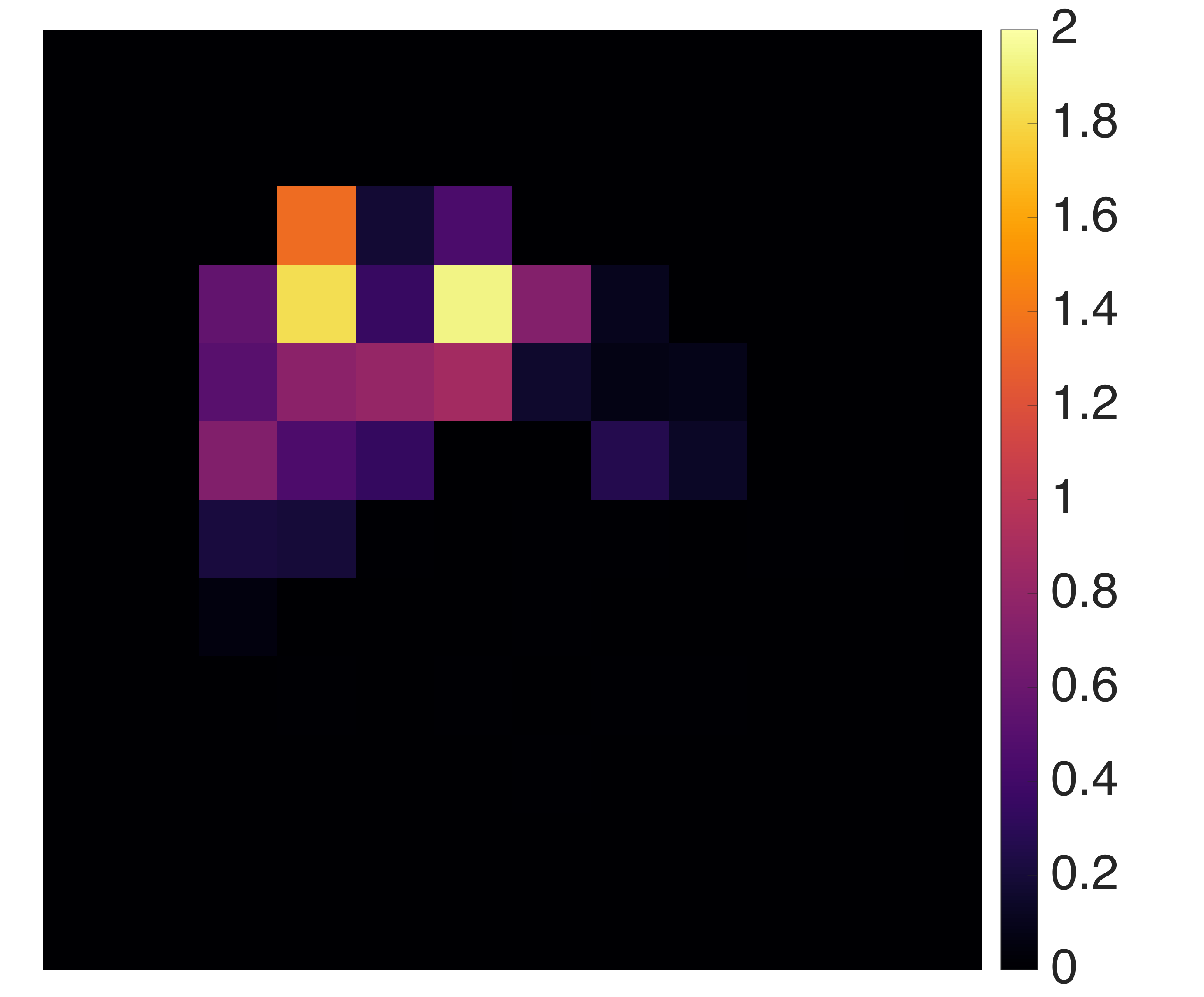} \label{fig:K2_kid}} 
\hfil
\subfloat[$k_{af}$]{\includegraphics[scale=0.1]{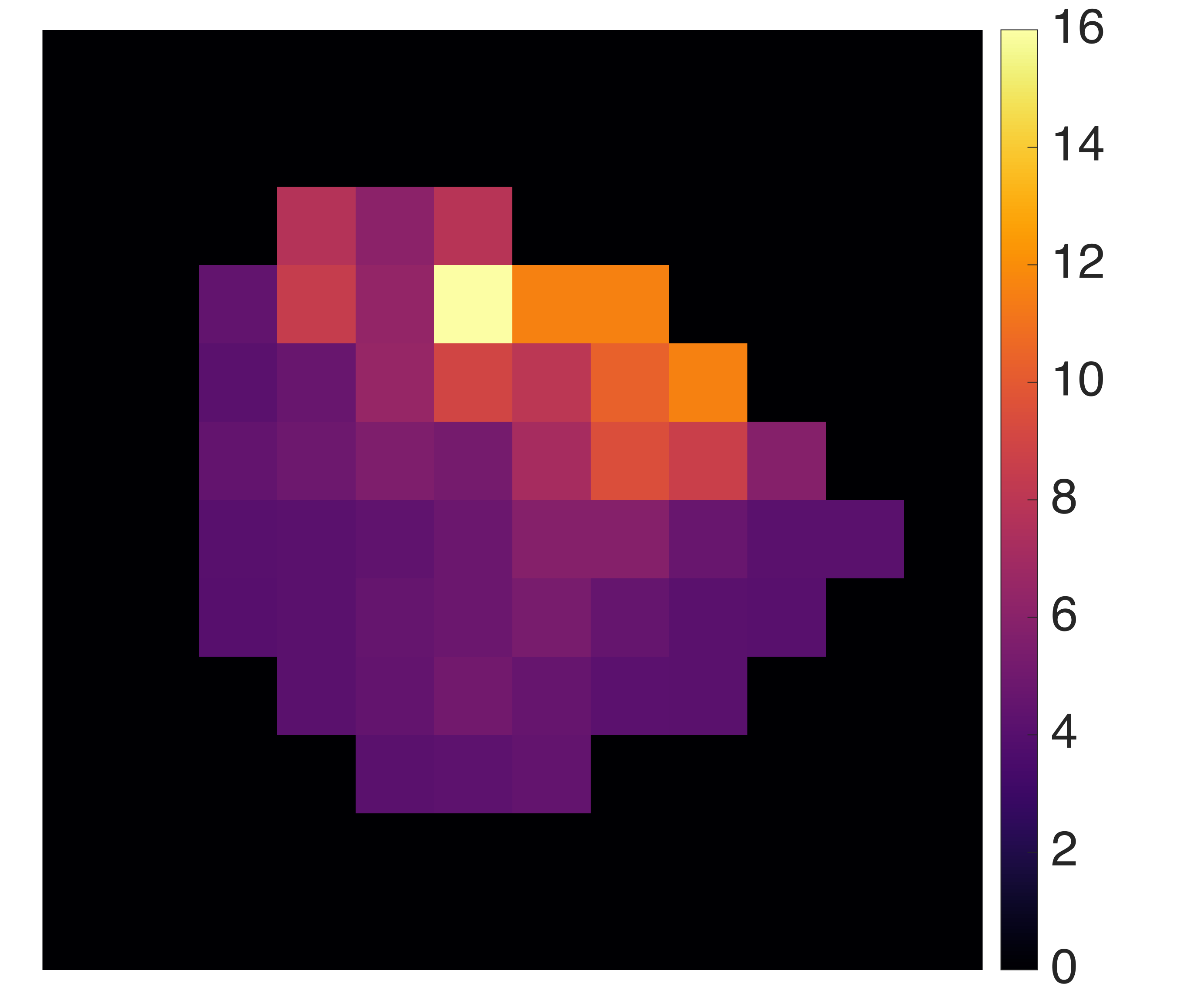} \label{fig:K3_kid}} 
\hfil
\subfloat[$k_{mf}$]{\includegraphics[scale=0.1]{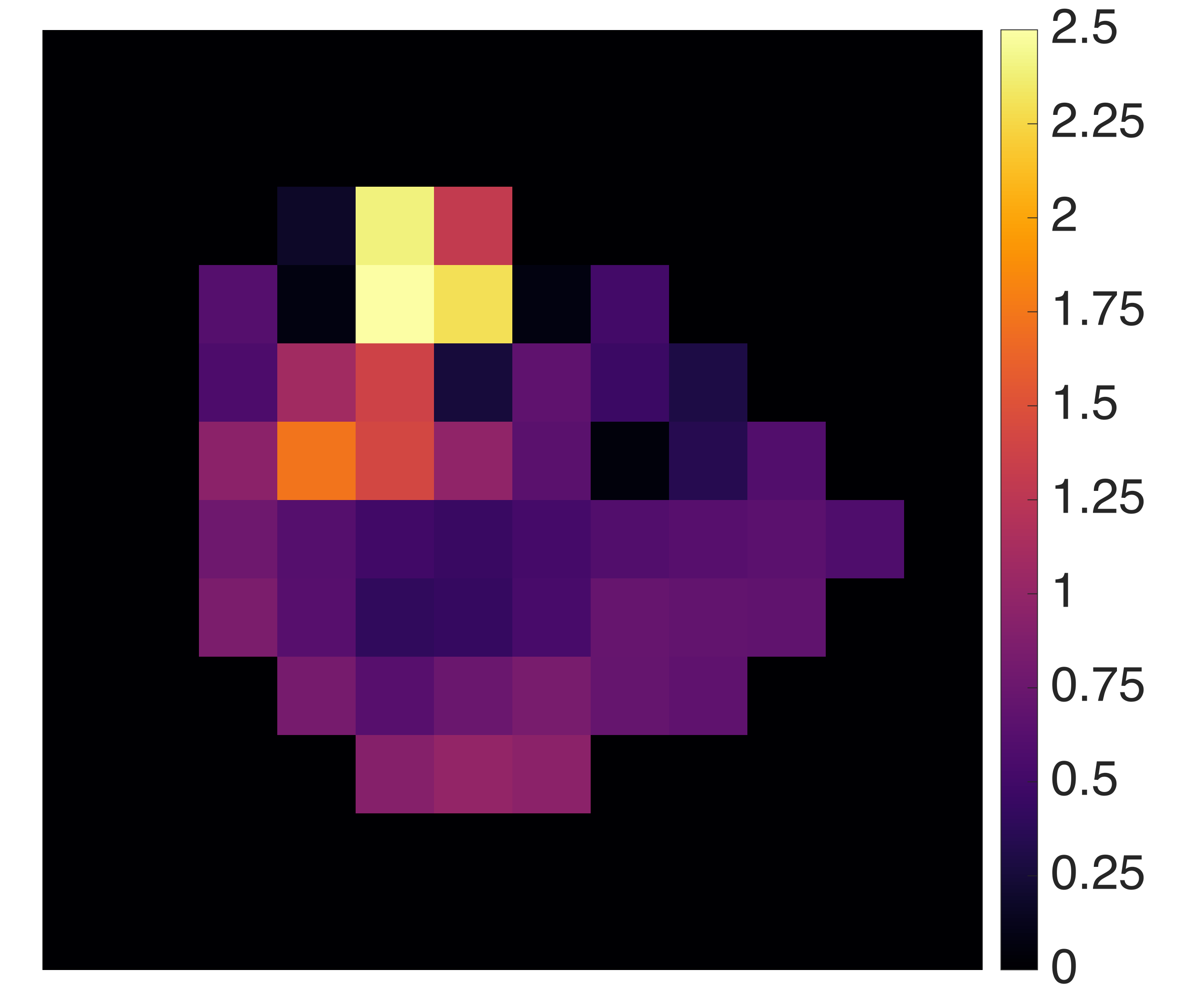} \label{fig:K4_kid}} 
\hfil
\subfloat[$k_{fm}$]{\includegraphics[scale=0.1]{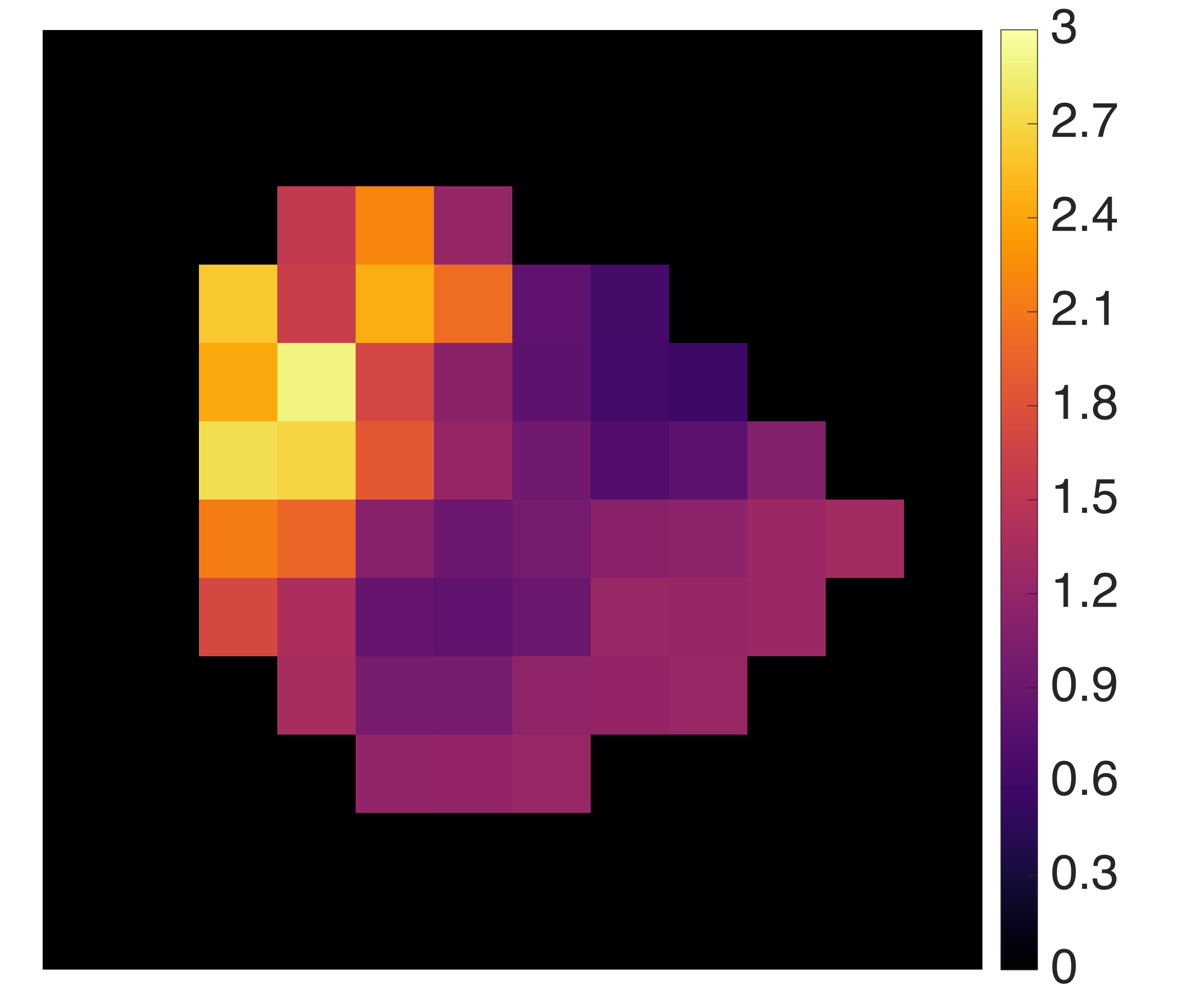} \label{fig:K5_kid}} 
\hfil
\subfloat[$k_{tm}$]{\includegraphics[scale=0.1]{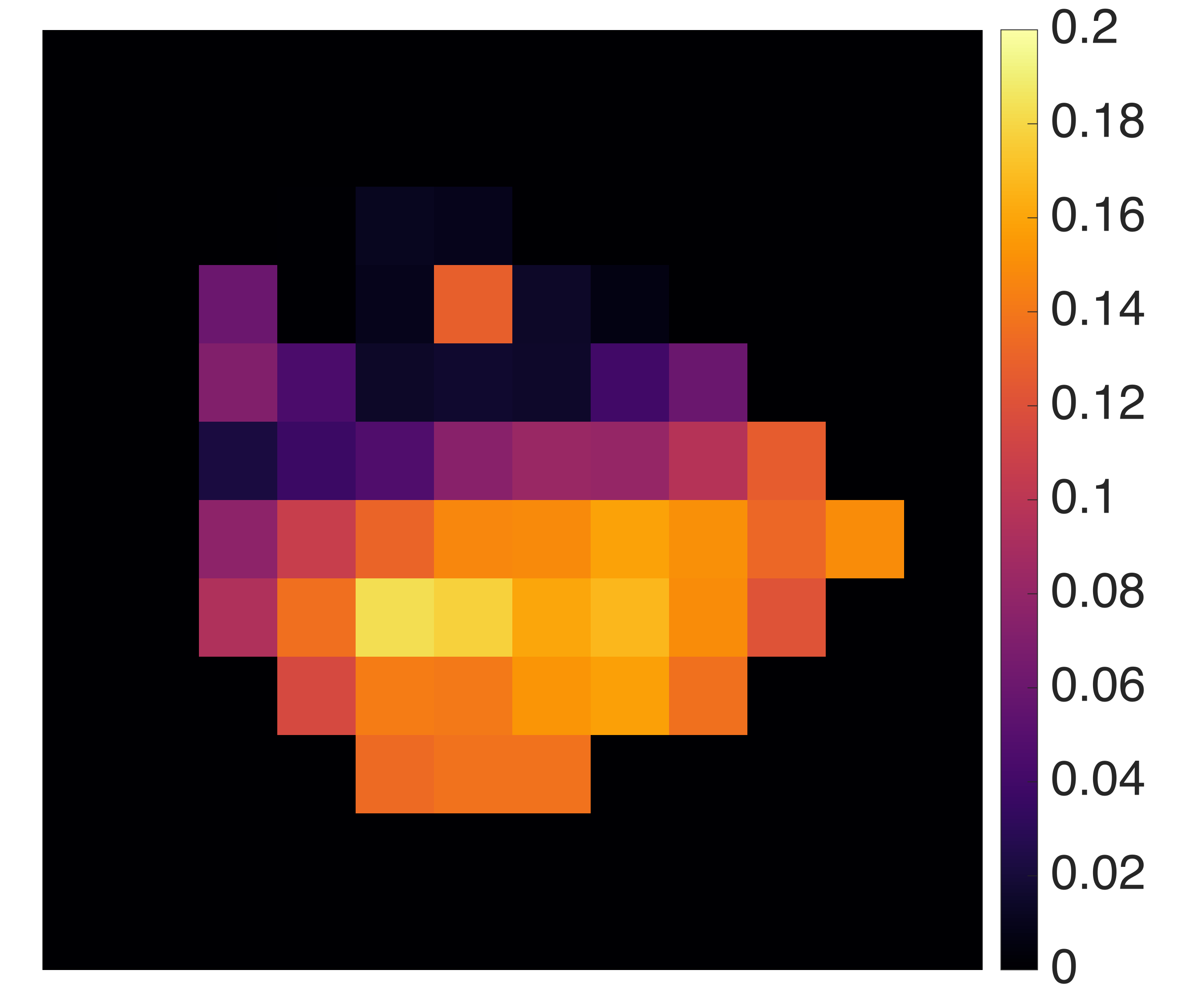} \label{fig:K6_kid}} 
\hfil
\subfloat[$k_{ut}$]{\includegraphics[scale=0.1]{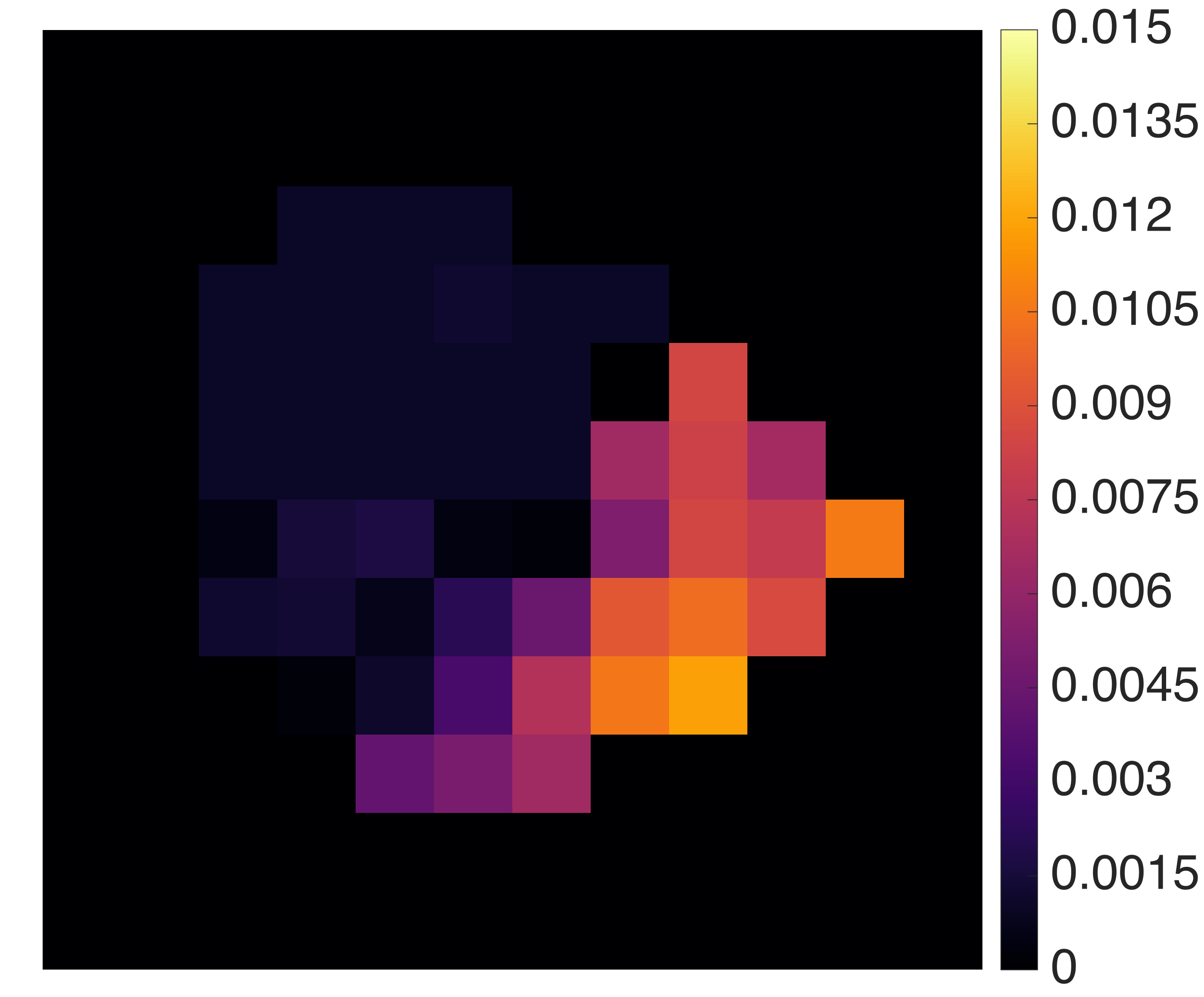} \label{fig:K7_kid}} 
\caption{Reconstructed parametric images for a selected PET slice of the renal system (right kidney).}
\label{fig:parametric_images_kidney}
\end{figure*}


Compartmental analysis \cite{Schmidt} identifies different functional compartments in the physiological system of interest, each one associated to a specific metabolic state of the tracer. The kinetics of the system, i.e. the time evolution of tracer concentration for each compartment, 
is formalized by a linear system of Ordinary Differential Equations (ODEs) with constant coefficients, expressing the conservation of tracer during the flow between compartments. The coefficients define the input/output rate of tracer for each compartment and represent the physiological parameters describing the metabolism of the system.
The reconstructed PET images of tracer concentration supply information on the tracer injected into the blood, mathematically modeled by the so-called Input Function (IF) of the compartmental system, and provide an estimate of the sum of the concentrations of the different compartments considered, at each time point of the acquisition. 
Given the forward model equation, i.e. the analytical solution of the system of ODEs, and the dynamic PET measurements of the tracer concentrations, the unknown kinetic parameters are estimated by means of an optimization scheme able to solve the non-linear ill-posed compartmental inverse problem. 
For the parametric imaging purpose, the exchange coefficients 
are not regarded as constant but as functions dependent on the spatial variable. Therefore, the compartmental inverse problem has to be solved for each pixel of the PET images.

We propose to determine the entire set of the so-called parametric images of tracer metabolism through a rather general parametric imaging method. 
We start from the set of reconstructed dynamic FDG-PET images of tracer concentration, select the tissue of interest, and formulate the compartmental model reliable for its functional description. For each dynamic PET image, our imaging method follows the steps described below.
\begin{itemize}
\item[{\bf 1.}] \emph{Gaussian smoothing.} In order to reduce the noise due to data acquisition, we apply a Gaussian smoothing filter through the convolution operation. 
\item[{\bf 2.}] \emph{Image segmentation.} To extract the region enclosing the organ of physiologic interest, we apply an heuristic image segmentation method based on a curve fitting process with a family of one dimensional gaussian function of variable variance.
\item[{\bf 3.}] \emph{Parameter estimation.} To estimate the kinetic parameters, we apply on each image pixel the regularized Gauss-Newton iterative algorithm \cite{Delbary_meth}.  
\item[{\bf 4.}] \emph{Parametric images.} Once we obtain the set of exchange coefficients of the model for each image pixel, we build up the parametric images.
\end{itemize}
We observe that item 2. can be upgraded by using more sophisticated image segmentation methods.

\section{Results}

We analyzed FDG-PET real data of murine models obtained by means of a microPET system (Albira, Carestream Health, Genova) currently operational at our lab.
The animals were studied after a fasting period of six hours, anesthetized and positioned on the bed of the microPET system. A dose of $3$-$4$ MBq of FDG was injected through the tail vein for an acquisition time of $40$ minutes. The dynamic PET images of tracer concentration (kBq/ml) were reconstructed using Maximum Likelihood Expectation Maximization (MLEM) with the following framing rate: $10 \times 15$s, $1 \times 22$s, $4 \times 30$s, $5 \times 60$s, $2 \times 150$s and $5 \times 300$s. We focused on the two-compartment catenary model describing the hepatic system \cite{Garbarino_liver} and the three-compartment non-catenary model representing the renal system \cite{Garbarino_kidney} (\figurename~\ref{fig:models}). We applied our parametric imaging method on different PET slices containing axial sections of the liver and axial sections of the kidneys. An example of the reconstructed parametric images obtained with our procedure for the hepatic and renal compartmental models are shown in \figurename~\ref{fig:parametric_images_liver} and \figurename~\ref{fig:parametric_images_kidney}, respectively.
The hepatic parametric images underly the homogeneity of the hepatic tissue, while the renal parametric images point out the different structures composing the kidney and characterizing the distinct functions of the organ, accordingly with the architecture of the respective compartmental model. 

\section{Conclusions}

This paper showed a novel physiology-based parametric imaging method, relying on the application of image processing algorithms and of an optimization scheme based on the regularized Gauss-Newton method. The experimental results proved the reliability of the method, whose main advantage is in its notable degree of generality, being it applicable, in principle, to models made of whatever number of compartments. 

\ifCLASSOPTIONcaptionsoff
  \newpage
\fi

\end{document}